\documentclass[prl,twocolumn,superscriptaddress,showpacs,amsmath,amssymb]{revtex4-1}

\usepackage{graphicx}
\usepackage{latexsym}
\usepackage{amsmath}
\usepackage{amssymb}
\usepackage{amsfonts}
\usepackage{color}
\usepackage{bm}
\usepackage{verbatim}
\usepackage[]{color}
\definecolor{red}{rgb}{1,0,0}
\usepackage{framed}
\definecolor{shadecolor}{RGB}{222,222,221}
\begin{document}

\title{Resistive state of SFS Josephson junctions in the presence of moving domain walls}

 \date{\today}
 
\author{D. S. Rabinovich}

\affiliation{Moscow Institute of Physics and Technology, Dolgoprudny, 141700 Russia}
\affiliation{Skolkovo Institute of Science and Technology, Skolkovo 143026, Russia}
\affiliation{Institute of Solid State Physics, Chernogolovka, Moscow
  reg., 142432 Russia}
  
\author{I. V. Bobkova}
\affiliation{Institute of Solid State Physics, Chernogolovka, Moscow
  reg., 142432 Russia}
\affiliation{Moscow Institute of Physics and Technology, Dolgoprudny, 141700 Russia}

\author{A. M. Bobkov}
\affiliation{Institute of Solid State Physics, Chernogolovka, Moscow reg., 142432 Russia}

\author{M.A.~Silaev}
 \affiliation{Department of
Physics and Nanoscience Center, University of Jyv\"askyl\"a, P.O.
Box 35 (YFL), FI-40014 University of Jyv\"askyl\"a, Finland}
\affiliation{Moscow Institute of Physics and Technology, Dolgoprudny, 141700 Russia}

 \begin{abstract}
 We describe resistive states of the system combining two types of orderings- superconducting and ferromagnetic one. It is shown that in the presence of magnetization dynamics such systems become inherently dissipative and in principle cannot sustain any amount of the superconducting current because of the voltage generated by the magnetization dynamics. 
 We calculate  generic current-voltage characteristics of a superconductor/ferromagnet/superconductor Josephson junction with an  unpinned domain wall and find the low-current resistance  associated with the domain wall motion. We suggest the finite slope of  Shapiro steps as the characteristic feature of the  regime with domain wall oscillations driven by the ac external current flowing through the junction.  
 \end{abstract}

  \maketitle

The ability to sustain dissipationless electric currents is assumed to be the defining property of superconducting state. However, this fundamental concept has been challenged by subsequent discovery of  type-II superconductors which can be driven into the mixed state characterized by the presence of Abrikosov vortices generated by the magnetic field \cite{Abrikosov1957}. The mixed state is generically resistive one since in the absence of additional constraints such as the geometrical confinement of the pinning potential the superconductor vortices start to move under the action of any external current\cite{Kim1965}. In such a  flux-flow regime vortex motion generates 
electric field which leads to the finite resistance and Ohmic losses\cite{Gorkov1975, Bardeen1965}.

In this Letter we point out one more fundamental mechanism which can drive superconducting system into the resistive state realized in the ideal situation for arbitrary small applied current.   
We find that the voltage can be generated in the superconductor/ferromagnet (S/F) systems   due to the interplay of two different order parameters known to produce many non-trivial effects\cite{BuzdinRMP2005,Bergeret2005,BergeretRMP2018,Eschrig2008,Houzet2007}.
The presence of superconducting condensate allows for the generation of dissipationless spin currents\cite{Eschrig2015} and spin torques to manipulate the  magnetic order parameter \cite{Waintal2002,Buzdin2008,Konshelle2009,
Teber2010,Holmqvist2012,Linder2011,Shukrinov2017,Chudnovsky2016,Braude2008,Nussinov2005,Zhu2004,Cai2010,Eschrig2008,Houzet2007,Bobkova2018,Rabinovich2018,Aikebaier2019}.
Here we point out that the  magnetization dynamics generated in this way by the supercurrent with necessity generates electric field and Ohmic losses in a way analogous to the Abrikosov vortex motion in the flux-flow regime. 
However, there is no complete analogy between these fundamental processes. In the case of magnetic system it is the dynamics of magnetic order parameter which generates electric field and Ohmic losses in the superconducting state  due to the Gilbert damping mechanism. 
The importance of this new resistive state for understanding the physics 
of non-equilibrium  superconducting/ferromagnet  systems
 motivates the present work.   

{ The sketch of the system under consideration is shown in Fig.~\ref{sketch}(a).
Its magnetic part consists of ferromagnetic strip and it conceptually similar to the domain wall (DW) racetrack memory proposal\cite{ParkinRacetrack1,ParkinRacetrack2}. The position of DWs  in the strip can be controlled by the normal current $j_N$, which can be applied along the strip.
 In addition two superconducting leads forming a Josephson junction are placed on top of a ferromagnetic strip.
 The Josephson current in such a geometry has been measured through CrO$_2$ half-metallic ferromagnet\cite{Singh2016} for the distance between superconducting electrodes much exceeding the typical DW size of $20$ nm. In such system the Josephson current is necessarily mediated by spin-triplet Cooper pairs \cite{Eschrig2008} which can pick up the Berry phase\cite{Bobkova2017} propagating through the non-coplanar  spin texture or in the presence of spin-orbit coupling (SOC). This leads to the possibility of spin transfer torques (STT) generated by the supercurrent\cite{Bobkova2018} so that  
  when a DW is located inside the interlayer region of the Josephson junction, it can be moved by the Josephson current $j$ applied between the superconducting leads. 

\begin{figure}[!tbh]
   \centerline{\includegraphics[clip=true,width=2.8in]{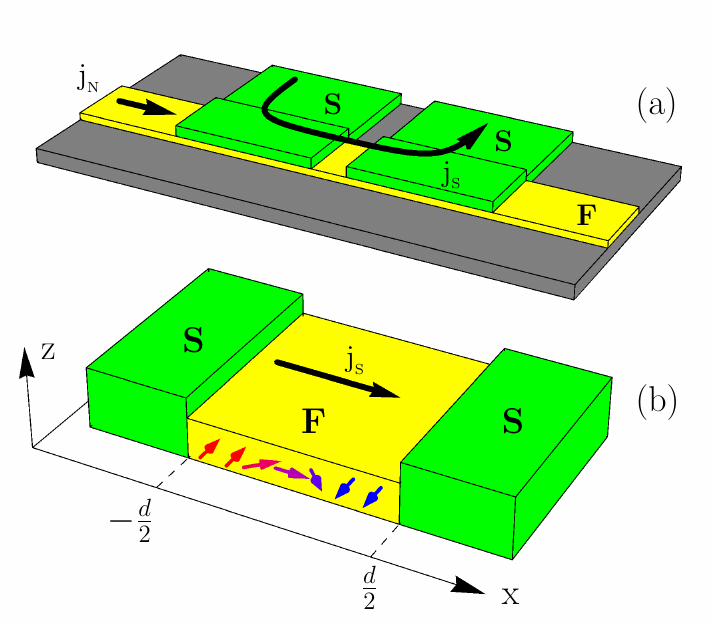}}
       \caption{(a) Sketch of the system under consideration. Superconducting electrodes forming a Josephson junction are placed on top of a ferromagnetic strip. The position of DWs in the strip can be controlled by the normal current $j_N$.
       (b) A simplified model of the Josephson junction region. A Neel-type DW is present in the interlayer.  The Josephson supercurrent flows in F region in $x$ direction. }
 \label{sketch}
 \end{figure}
 %

 
  As a reciprocal effect to the STT a gauge spin-dependent vector potential appears in the local spin basis due to SOC \cite{Frohlich1993,Rebei2006,Jin2006,Jin2006_2,Bernevig2006,Hatano2007,Leurs2009,Tokatly2008,Bergeret2013}. It produces an anomalous phase shift and, in case if the magnetization depends on time, it also produces an electromotive force\cite{Kim2012,Tatara2013,Yamane2013}. This situation is the focus of our present study. The electromotive force should be compensated by the voltage induced at the junction. It is this voltage that maintains the DW motion, compensating the dissipation power occurring due to Gilbert damping by the work done by a power source, as it is shown below.
 
} 

{\it The model.} 
    Fig.~\ref{sketch}(b) illustrates the simplified model of the SFS Josephson junction region, which we consider in our calculations. We assume that there is a Neel-type DW inside the F interlayer.  The Rashba SOC is present in F due to structural or internal inversion symmetry breaking. The Josephson supercurrent 
  which flows in F  along the $x$-direction generates a torque on the DW 
  \cite{Bobkova2018} 
  consisting of the adiabatic STT \cite{Slonczewski1996,Tatara2004,Koyama2011} 
 and spin-orbit torques\cite{Miron2010,Gambardella2011}.
 Under these conditions 
 DW motion is caused even by very small currents if pinning effects are neglected. 
 We  neglect the nonadiabatic  STT\cite{Zhang2004} assuming that the most part of the voltage is dropped at the interfaces and the quasiparticle nonequilibrium in the interlayer is small enough.

In the considered S/F/S junction the  coupled dynamics of magnetization 
$\bm M$
and Josephson phase difference
 $\varphi$ is determined by the following 
 closed set of equations 
\begin{align} \label{josephson_modified} 
  & j=j_{c}\sin \left(\varphi-\varphi_0\{\bm M\} \right) + \frac{\dot \varphi-\dot \varphi_0 \{\bm M\}  }{2eRS}.
 \\ \label{LLG}
 & \frac{\partial\bm M}{\partial t} = -\gamma \bm M \times \bm H_{eff} + \frac{\alpha}{M} \bm M \times \frac{\partial\bm M}{\partial t} + \bm T,
\end{align}
 Eq.(\ref{josephson_modified}) represents the 
 non-equilibrium current-phase relation (CPR)
 generalizing 
 resistively shunted Josephson junction (RSJ) model. 
 This relation is written in a gauge-invariant 
 form amended to include the anomalous phase shift
 \cite{Krive2004, Braude2007, Asano2007, Reynoso2008, Buzdin2008, Tanaka2009,
Grein2009, Zazunov2009, Liu2010,Malshukov2010,Alidoust2013,Brunetti2013,Yokoyama2014,
Kulagina2014,Moor2015_1,Moor2015_2,Bergeret2015,Campagnano2015,Mironov2015,Konschelle2015,
Kuzmanovski2016,Zyuzin2016,Bobkova2016,Silaev2017,Bobkova2017,Rabinovich2018,Szombati2016,Murani2017,Assouline2019,Meng2019}
 $\varphi_0\{\bm M\}$ defined by 
 SOC and magnetic texture.
  For  strong ferromagnets only spin-triplet pairs can penetrate into F. Then the transport can be calculated in the local spin basis for spin-up and spin-down Fermi surfaces separately with effective  U(1) spin-dependent gauge field $\bm Z$
 that yields\cite{Bobkova2017}
\begin{eqnarray}
\varphi_0 \{\bm M \} = -2\int \limits_{-d/2}^{d/2} Z_x(x,t)dx.
\label{phi_0_general}
\end{eqnarray}
 where
 $\bm Z= \bm Z^{m} + \bm Z^{so}$. Here 
 $Z^{m}_i = -i {\rm Tr} \Bigl( \hat \sigma_z \hat U^\dagger \partial_i \hat U \Bigr)/2 $ is the texture-induced part, where $\hat U(\bm r, t)$ is the
time- and space-dependent unitary $2 \times 2$ matrix that rotates
the spin quantization axis $z$ to the local frame determined by
the exchange field.

The term $Z^{so}_j = (M_i B_{ij})/M$ appears due to SOC, where $B_{ij}$ is the constant tensor coefficient describing the linear spin-orbit coupling of the general form $\hat H_{so} = \sigma_i B_{ij} p_j/m$. Here we assume that $\hat H_{so}$ is of Rashba type: $\hat H_{so} = (B_R/m) (\sigma_x p_y - \sigma_y p_x)$. $\bm Z^m$ is nonzero only  for noncoplanar magnetic structures 
and in our case $\bm Z^m = 0$
\footnote{It is worth to mention that for an in-plane exchange field as is shown in Fig.~\ref{sketch}(b) the Rashba SOC by itself does not generate long-range triplet correlations at S/F interfaces. Therefore, to produce them spin-active layers should be added to the system at the S/F interfaces. However, this problem is widely studied in the literature\cite{Bergeret2005,Eschrig2008,Houzet2007} and is not discussed here.}.
The electromotive force can also occur due to noncomplanarity of the moving DW or the presence of the nonadiabatic (anti-damping) torque \cite{Volovik1987,Barnes2007,Saslow2007,Duine2008,Tserkovnyak2008,Zhang2009}. However, for the case of Rashba SOC and the Neel DW, presented in Fig.~\ref{sketch}(b), the moving DW remains coplanar.

In fact, Eq.~(\ref{josephson_modified}) is quite general and is applicable to a wide class of Josephson systems exhibiting anomalous phase shift. We derive it  microscopically in Supplementary material for the case of a strong ferromagnetic interlayer\cite{Bobkova2017}.
Besides that
in contrast to the previously used gauge non-invariant formulations\cite
{Konshelle2009,Shukrinov2017} Eq.(\ref{josephson_modified}) describes the normal spin-galvanic effects when $j_c=0$ such as the
 electromotive force and charge current generated in the ferromagnet due to the  time
derivative of the Berry phase
\cite{Volovik1987, Barnes2007, Saslow2007, Duine2008,
Tserkovnyak2008,Schulz2012,Nagaosa2013}.  The analogous equation is also valid for a more general nonsinusoidal CPR.

 
 The magnetization dynamics driven by the spin-polarized supercurrent is described by LLG Eq.(\ref{LLG}) where $\alpha$ is the Gilbert damping constant, $\gamma$ is the gyromagnetic ratio. 
 The last term in  Eq.(\ref{LLG})
 is the current-induced spin torque  
 $\bm T = (\gamma/M)(\bm J_s \bm \nabla)\bm M + (2  \gamma /M)(\bm M \times \bm B_j) J_{s,j}$. 
 The first term here is the 
  adiabatic spin-transfer torque generated by the 
  spin current $\bm J_s$.
  The second term is the spin-orbit induced torque determined by the spin vector $\bm B_j = (B_{xj},B_{yj},B_{zj})$ corresponding to the $j$-th spatial component of the SOC tensor $B_{ij}$.  Below we assume $R_{\uparrow} \ll R_{\downarrow}$ and neglect for simplicity  the spin-down contribution to the current. In this case $\bm J_s \approx \bm j/2e$. 

 {\it DW motion.} It is convenient to parametrize the magnetization as 
  $\bm  M=  M(\sin \theta \sin \delta, \cos \theta, \sin \theta \cos \delta )$,
 where the  both angles depend on $(x,t)$.
 At zero applied supercurrent the equilibrium shape of the DW is given by $\delta = \pi/2$ and
{ \begin{equation} \label{theta}
 \cos\theta = - \tanh [(x-x_0)\pi/d_W],
  \end{equation}
 where $d_W = \pi \sqrt{A_{ex}/K}$ is the DW width.} Here it is assumed that $K>0$ and $K_\perp >0$ are the anisotropy constants for the easy and hard axes, respectively and $A_{ex}$ is the constant describing the inhomogeneous part of the exchange energy. The effective magnetic field $\bm H_{eff} = (1/M^2) (K M_y \bm y - K_\perp M_z \bm z + A_{ex} \partial_x^2 \bm M)$.
 
 
 For dealing with the SOC-induced torque it is convenient to define the dimensionless SOC constant   $\beta = -2B_R d_W/\pi$. For small applied supercurrents $j \ll d_W M e/ (\pi t_d \mu_B |\alpha - \beta|)$ the DW moves as a coplanar object corresponding to  $\theta (x,t)$ defined by Eq.~(\ref{theta}) with $x_0(t) = \int \limits_{t_0}^t v(t')dt'$ and $\delta \equiv \delta(t)$. The exact solution for $v(t)$ following from Eq.~(\ref{LLG}) for the situation when the electric current is switched on at $t=0$ is presented in the Supplementary material.
 On a characteristic time scale $t_d = (1+\alpha^2)M/(\alpha \gamma K_\perp)$
 the DW velocity  reaches its stationary value
 \begin{align}  \label{velocity_steady}
 v_{st}= - u\beta/\alpha,
 \end{align}
where $u = \gamma J_s /M$. 


In the considered case of Neel DW and Rashba SOC  $Z_x^{so} = (\pi \beta M_y)/(2d_W M)$ which according to (\ref{phi_0_general}) yields the anomalous phase shift:
 \begin{eqnarray}
\varphi_0 (t) \approx -2 \pi \beta x_0(t) /d_W.
\label{varphi_0_approx}
\end{eqnarray}
Our consideration is strictly applicable if $|d/2 \pm x_0| \gg d_W$, that is if the DW is not close to the S/F interface. 

\begin{figure}[!tbh]
 \begin{minipage}[b]{\linewidth}
   \centerline{\includegraphics[clip=true,width=3.2in]{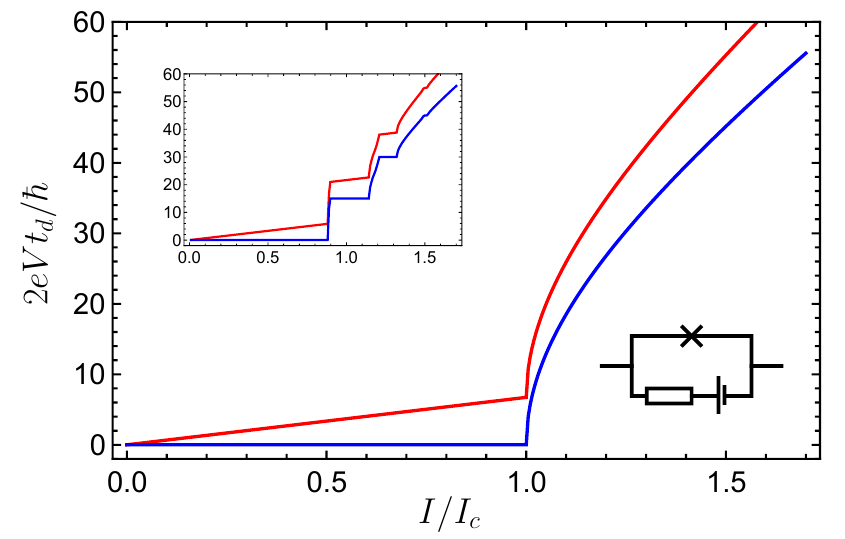}}
   \end{minipage}
      \caption{IV-characteristics of the SFS junction with a DW at rest (blue) and a moving DW (red). $\beta = 1$, $\alpha = 0.1$, $eKd_W/(\pi j_c)=5$, $t_d=40t_J$, where $t_J=1/2eRI_c$. Upper-left insert: Shapiro steps for $I(t)=I + 0.3I_c \cos \omega t$, $\omega = 15 t_d^{-1}$. Axes labeling is the same as in the main figure. Bottom-right insert: the equivalent circuit scheme of the junction.}
 \label{dc_j}
 \end{figure}

{\it Resistive state.}
Suppose that we apply a constant electric current $I=jS$  (here $S$ is the junction area) to the Josephson junction and consider a steady motion of the DW across the junction with a constant velocity defined by Eq.~(\ref{velocity_steady}). In this case Eq.~(\ref{josephson_modified}) can be easily solved and the time-averaged voltage induced at the junction is
{ \begin{eqnarray}
\overline{V(t)} = RS\sqrt{j^2-j_c^2}+\dfrac{\pi \beta^2 u}{e\alpha d_W},
\label{V_averaged}
\end{eqnarray}}
where the first term represents the well-known Josephson voltage, which is generated at $j>j_c$. The second term $V_M $ is nonzero as at $j>j_c$, so as at $j<j_c$ and reflects the fact that the Josephson junction is in the resistive state if the DW is moved by the current. The corresponding IV- characteristics of the junction are shown in Fig.~\ref{dc_j}. In principle, in small-area or point Josephson junctions with large resistance the capacitance or inductance shunting can also lead to the finite slope of the supercurrent branch \cite{Stewart1968,McCumber1968,Kautz1990}. At the same time, the experimentally realized Josephson junctions via metallic ferromagnets \cite{Singh2016,Golod2019} practically do not demonstrate noticeable slopes of the supercurrent branches. However, even if a finite slope due to the interaction with environment is present, it can be distinguished experimentally from the effect discussed in our manuscript by comparing IV-characteristics of the same junction in the presence and in the absence of the domain wall. It is also worth noting that Eq.~(\ref{josephson_modified})  does not yield a ratchet potential for the Josephson phase because the anomalous phase shift can be compensated by the change of origin. Therefore, our model yields no rectification effects typical for  asymmetric Josephson systems \cite{LJJRatchet1, LJJRatchet2, LJJRatchet3, JJarrayRatchet, asymSQUIDRatchet1,asymSQUIDRatchet2, 3JSQUIDRatchet1, 3JSQUIDRatchet2, AnnularJJRatchet, asymDriveJJRatchet1, asymDriveJJRatchet2, AbrikosovVortJJRatchet} and the IV-characteristics remains purely antisymmetric with respect to the current reversal. 

{ For numerical estimates of $V_M$ we take $\alpha=0.01$, $d_W = 60 nm$, $u \approx 1 m/s$, what corresponds to the maximal Josephson current density\cite{Singh2016}  through the $\rm {CrO}_2 $ nanowire $j_c \sim 10^9$ ${A/m}^2$. The dimensionless SOC constant $\beta$ can vary in wide limits. Having in mind that experimentally our predictions can be realized, for example, for hybrid interlayers consisting of a ferromagnet/heavy metal bilayers, $\beta = 1-10$ considering that the SOC $\alpha_R=B_R/m$ ranges from $3 \times 10^{-11}$ to $3 \times 10^{-10} eV m$ at interfaces of heavy-metal systems \cite{Ast2007}. Then we can obtain $V_M|_{j=j_c}$ up to $10^{-5} - 10^{-3}V.$ }

The resistance of the junction at $j<j_c$ caused by the DW motion is given by
{ \begin{eqnarray}
R_{DW} = \left(\frac{\partial V}{\partial I}\right)_{I<I_c} = \frac{\pi \gamma \beta^2\hbar}{2e^2 S\alpha d_W M},
\label{R_averaged}
\end{eqnarray}}
It is interesting that according to Eq.~(\ref{R_averaged}) $R_{DW}$ per unit area does not depend on the Josephson junction parameters, such as $j_c$ and $R$, and is determined only by the characteristics of the magnetic subsystem. It can be naturally understood if we take into account that in this case the work done by a power source is exactly equal to the energy losses due to the Gilbert damping. Indeed, the dissipation power due to Gilbert damping can be calculated as \cite{Tserkovnyak2009}
\begin{eqnarray}
P_G = \frac{\alpha}{\gamma M}\int dx \Bigl( \frac{d \bm M}{d t} \Bigr)^2
\label{G_losses}
\end{eqnarray}
For the  stationary DW motion 
described by Eq.~(\ref{theta}) with $\dot x_0 = v_{st}$ 
we get  $P_G =j (\pi u \beta^2/e \alpha d_W)$,
which exactly coincides with the power $jV$ provided by the source. 

In the regime $j<j_c$ the normal current through the Josephson junction is zero in spite of nonzero voltage generated at the junction. This follows directly from Eq.~(\ref{josephson_modified}) because 
 for $j<j_c$ it has the solution $\dot \varphi(t)=\dot \varphi_0(t)$. The equivalent circuit scheme of the junction is presented in the insert to Fig.~\ref{dc_j}. The voltage is compensated by the electromotive force induced in the junction by the emergent electric field $(\hbar/e) \dot{\bm Z}^{so}$. 

If there are $n$ DWs inside the junction, then under the assumptions above $R_{DW}$ expressed by Eq.~(\ref{R_averaged}) is multiplied by $n$. 
 If the Josephson junction is driven by an ac component of the voltage or current having the frequency $\omega$, then the dependence $V(I)$ manifest horizontal steps at $V_k = k \omega/2e$, which are known as the Shapiro steps\cite{Shapiro1963,Grimes1968}. If a moving DW is present in the junction the Shapiro steps acquire a nonzero slope, which is determined by Eq.~(\ref{R_averaged}). The reason  is that in this case the oscillation frequency of the Josephson current  is determined by $\overline{\dot \varphi - \dot \varphi_0}$ and does not coincide with $2eV$ anymore. The Shapiro steps occur just when the oscillation frequency of the Josephson current equals to multiple integer of the external frequency. The $IV$-characteristic demonstrating the inclined Shapiro steps is shown in the insert to Fig.~\ref{dc_j}.   

In real setups 
the time of the DW movement through the junction is limited by the finite junction length: $t_{DW} \approx d/v_{st} = (\alpha/\beta) (d/u)$. Therefore, the voltage should be averaged over  $t<t_{DW}$. 
Although   experimental data  on the DW motion in Josephson junctions has not been available yet, 
for the estimates we take $d=0.5 \times 10^{-6}$ m and $u  \approx 1$ m/s. Then $t_{DW} \geq 0.5 (\alpha/\beta) \times 10^{-6}$s. For other experiments, where the Josephson current carried by equal-spin triplet correlations was reported \cite{Khaire2010,Robinson2010}, this time can be several orders of magnitude higher due to much less values of the critical current density.  

The IV-charactiristic presented in Fig.~\ref{dc_j} was obtained under the assumption of  a steady DW motion. In fact, $V(t)$ is driven by $\dot \varphi_0 \sim v(t)$. For a step-like applied electric current $V(t)$ saturates exponentially at the characteristic time $t_d$ except for the short Josephson pulses (see below). Therefore, in order to be able to measure the resistance expressed by Eq.~(\ref{R_averaged}) it is important to have $t_{DW}>t_{d}$. For estimations of $t_d$ we use the material parameters of the CrO$_2$ nanostructures \cite{Zou2007,Singh2016}. Taking the saturation magnetization $M = 4.75\times 10^5\; A/m$, $K = 1.43 \times 10^5 erg/cm^3$ and $K_\perp = 4 \pi M^2$, $\alpha=0.01$ we obtain $t_d \approx  10^{-9} $s.  
  Consequently, the ratio $t_{DW}/t_d > 1/\beta $ and for not very large values of the SOC constant $\beta \lesssim 1$ the condition $t_{DW}>t_{d}$ is realistic.
  
\begin{figure}[!tbh]
   \centerline{\includegraphics[clip=true,width=3.0in]{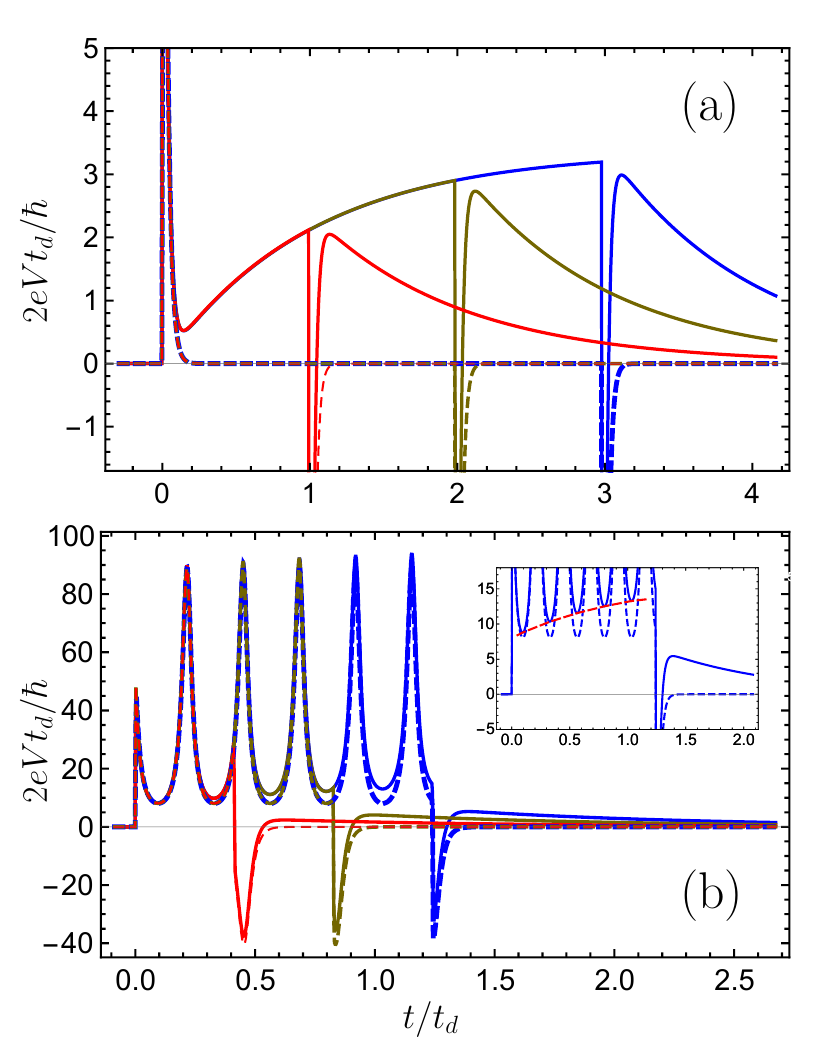}}
        \caption{$V(t)$ for rectangular current impulses. Different curves correspond to different impulse periods $T$; For all the panels the solid lines correspond to $\beta = 1$ (the anomalous phase due to the DW motion is nonzero) and the dashed lines are for $\beta = 0$ (the anomalous phase shift is zero). (a) $j=0.5j_c$, $T=3t_d$ (blue), $T=2t_d$ (yellow), $T=t_d$ (red); (b) $j=1.2j_c$, $T=1.25t_d$ (blue), $T=0.83t_d$ (yellow), $T=0.42t_d$ (red). Insert: $j=1.2j_c$, $T=1.25t_d$ (the part of the main panel on a large scale); For all the curves   $\alpha = 0.1$, $eKd_W/(\pi j_c)=5$, $t_d= 40 t_J$.}
 \label{pulse_j}
 \end{figure}
 
 In practice DW motion can be
  induced by large current pulses. 
 For short pulses
 $j(t)=j\theta(t)\theta(T-t)$  with $T<t_{DW}$, the DW does not go out of the junction during the impulse time. The exact expression for the DW velocity $v(t)$ is to be found from the LLG equation and is calculated in the Supplementary material. The resulting voltage signal consists of two parts of different physical origin. The first part is the purely Josephson response with the characteristic time $t_J=1/2eRI_c$ and the other part is of purely magnetic origin, has time scale $t_d$ and vanishes if there is no moving DW in the junction. 
Taking for estimates of $t_J$ the material parameters of the CrO$_2$ nanostructures $j_c \sim 10^9 {\rm A/m^2}$, $R \sim 0.3 - 1.5 {\rm \Omega}$, $S=7.5 \times 10^{-14} {\rm m^2}$  we obtain $t_J = 0.3 \times 10^{-11}- 1.5 \times 10^{-11} {\rm s}$. According to this estimate $t_J \ll t_d$. Then the Josephson voltage signal should decay much faster than the DW signal. 

The resulting voltage signals  for $j<j_c$ 
are shown in Fig.~\ref{pulse_j}a.
In this regime the typical $V(t)$ curve consists of an initial sharp Josephson voltage impulse decaying at $t \sim t_J$, a final sharp impulse of the same nature and a gradual voltage increase and decrease of purely magnetic origin, which takes the form $ V(t) = - \pi \beta v(t)/e d_W$.
 %
  %
%
The signal saturates at time  $t_d$ to the voltage defined by the steady DW motion velocity.
In the absence of DW 
there is no generated voltage between the initial and final Josephson pulses as shown by dashed curves in Fig.~\ref{pulse_j}(a).

The regime  $j>j_c$ is characterized by the voltage signal  Josephson oscillations during the impulse time as shown in Fig.~\ref{pulse_j}(b). Nevertheless, the gradual increase of the voltage due to the DW motion is also present. It results in the   increasing difference between the solid and the dashed curves minima as marked by a red dashed line in the insert.  

To conclude we have generalized the RSJ equation to describe the new resistive state generated by
magnetization dynamics in the interlayer of a S/F/S junction.  
Taking into account the emergent 
vector potential originated from SOC and/or magnetization texture we obtained the gauge-invariant system of coupled equations which governs the dynamics of magnetization and superconducting phase.  
Using this model we have shown that in the presence of magnetization dynamics the Josephson junction is in the resistive state even at $j<j_c$. In this regime the junction can be used for electrical detection of the dynamics. Experimentally DW motion inside the Josephson junction can also be observed through the nonzero slope of Shapiro steps.  
 %

\begin{acknowledgments} 
The research of I.V.B and A.M.B has been carried out within the state task of ISSP RAS with the support by RFBR grant 19-02-00466. D.S.R acknowledges the support by RFBR grants 19-02-00466 and 19-02-00898. I.V.B and D.S. R also acknowledge the financial support by Foundation for the Advancement of Theoretical Physics and Mathematics “BASIS”. The work of M.A.S was supported by the Academy of Finland (Project No. 297439) and  Russian Science Foundation, Grant No. 19-19-00594.
\end{acknowledgments} 

\section*{Supplementary material}

\subsection*{Derivation of the modified RSJ equation}

For the system under consideration the interlayer is assumed to be made of a strong ferromagnet and can be described by the Green's function $\check g_\sigma$, defined at spin-up and spin-down Fermi surfaces separately, which is a $4 \times 4$ matrix in the particle-hole and Keldysh spaces, but is a scalar in spin space. In the local reference frame, where the
spin quantization axis is aligned with the local direction of the
exchange field in the ferromagnet it obeys the following generalized Eilenberger-Keldysh equation \cite{Bobkova2017,Bobkova2018}
\begin{equation} \label{Eq:UsadelEquation}
 -iD_{\sigma} \hat\partial_{\bm R}(
 \hat g_{\sigma} \otimes \hat\partial_{\bm R} \hat g_{\sigma}) +
 [ \varepsilon \hat\tau_3, \hat g_{\sigma} ]_\otimes =0,
 \end{equation}
 where $\tau_i$ are Pauli matrices in the particle-hole space,
 the $\otimes$-product is defined as $\hat A(\varepsilon,t) \otimes \hat B(\varepsilon,t) = \exp[(i/2)(\partial_{\varepsilon_A}\partial_{t_B}-\partial_{\varepsilon_B}\partial_{t_A})]\hat A(\varepsilon,t) \hat B(\varepsilon,t)$. The operator $\hat \partial_{\bm R}$ means the covariant derivative $\hat \partial_{\bm R} = \partial_{\bm R} + i \sigma [Z\tau_3, ...]_\otimes$ with $[\hat A, \hat B]_\otimes = \hat A \otimes \hat B - \hat B \otimes \hat A$. $\bm Z$ is the U(1) gauge field
which is added to the usual electromagnetic vector potential
$\bm A$ with the opposite effective charges for spin-up and spin-down
Fermi-surfaces\cite{Bobkova2017}. In general, the spin-dependent gauge field is given by the superposition of two terms
 $\bm Z= \bm Z^{m} + \bm Z^{so}$. 
 $Z^{m}_i = -i {\rm Tr} \Bigl( \hat \sigma_z \hat U^\dagger \partial_i \hat U \Bigr)/2 $ is the texture-induced part, where $\hat U(\bm r, t)$ is the
time- and space-dependent unitary $2 \times 2$ matrix that rotates
the spin quantization axis $z$ to the local frame determined by
the exchange field. 

The term $Z^{so}_j = (M_i B_{ij})/M$ appears due to  SOC, where $B_{ij}$ is the constant tensor coefficient describing the linear SOC of the general form $\hat H_{so} = \sigma_i B_{ij} p_j/m$. Here we assume that $\hat H_{so}$ is of Rashba type: $\hat H_{so} = (B_R/m) (\sigma_x p_y - \sigma_y p_x)$. $\bm Z^m$ only nonzero for noncoplanar magnetic structures. In this case the DW is moved as a plane object, consequently $\bm Z^m = 0$. The spin-orbit part is the only source of the spin-dependent gauge field in our case.

The general expression for the electric current inside the strong ferromagnet can be written as a sum over two spin subbands \cite{Bobkova2018}:
\begin{eqnarray}
j=\frac{e}{8} \sum \limits_\sigma \nu_\sigma D_\sigma {\rm Tr}_2 \int \bigl[ \tau_3 \hat g_\sigma \otimes \hat \partial_x \hat g_\sigma \bigr]^K d \varepsilon, 
\label{current_general}
\end{eqnarray}
where $\nu_\sigma$ and $D_\sigma$ are the density of states at the Fermi level and the diffusion constant for a given spin subband. $[...]^K$ means the Keldysh component of the corresponding matrix. 

In general, the current can be viewed as a sum of supercurrent and a normal current, carried by quasiparticles. Here we consider  high-temperature limit $\Delta(T) \ll T_c$ and linear response theory. We only keep the leading order terms in expressions for supercurrent and for normal current. This means that the supercurrent is calculated with the equilibrium distribution function neglecting nonequilibrium corrections and the normal current is calculated in the linear order with respect to the potential drop $V$ and for $\Delta \to 0$. Therefore, we neglect the terms of the order $(\Delta^2/T_c^2)(eV/T_c)$ upon the current calculation. The supercurrent in the framework of this approximation has been already calculated \cite{Bobkova2017} and is represented by the first term of Eq.~(\ref{josephson_modified}) of the main text. Here we are only interested in the expression for the normal current in the presence of the emergent electric field induced by the SO coupling in the weak link of the junction. At $\Delta \to 0$ $\hat g_\sigma^{R,A} = \pm \tau_3$ and $\hat g_\sigma^K = 2 \tau_3 \hat \varphi_\sigma$, where $\hat \varphi_\sigma = \varphi_\sigma^{e}(1+\tau_3)/2 + \varphi_\sigma^{h}(1-\tau_3)/2$ is the quasiparticle distribution function for a given spin subband. Eq.~(\ref{current_general}) is simplified as follows:
\begin{eqnarray}
j=\frac{e}{2} \sum \limits_\sigma \nu_\sigma D_\sigma \int \hat \partial_x \varphi_\sigma^e d \varepsilon. 
\label{current_normal}
\end{eqnarray}
\begin{eqnarray}
\hat \partial_x \varphi_\sigma^e \approx \partial_x \varphi_\sigma - \frac{\sigma}{2}(\partial_{\varepsilon_1}\partial_{t_2}-\partial_{\varepsilon_2}\partial_{t_1}) \times ~~~~\nonumber \\
\bigl[ Z_x^{so}(t_1)\varphi_\sigma(\varepsilon_2) - Z_x^{so}(t_2)\varphi_\sigma(\varepsilon_1) \bigr]\Bigl |_{\varepsilon_1 = \varepsilon_2 = \varepsilon; t_1 = t_2 = t} =\nonumber \\  \partial_x \varphi_\sigma^e + \sigma \partial_\varepsilon \varphi_\sigma^e \dot Z_x^{so} ~~~~~~~~~~~~~
\label{varphi_derivative}
\end{eqnarray}
Assuming that the quasiparticle nonequilibrium in the interlayer can be described by the Fermi distribution function with a spatially-dependent electric potential $\varphi_\sigma = \tanh [(\varepsilon - e V_\sigma(x))/2T]$ we can further simplify Eq.~(\ref{varphi_derivative}):
\begin{eqnarray}
\hat \partial_x \varphi_\sigma^e \approx  \partial_\varepsilon \varphi_\sigma^e \bigl(-e \partial_x V_\sigma + \sigma \dot Z_x^{so}\bigr).
\label{varphi_derivative_2}
\end{eqnarray}
Substituting Eq.~(\ref{varphi_derivative_2}) into Eq.~(\ref{current_normal}) one can obtain
\begin{eqnarray}
j=-e \sum \limits_\sigma \nu_\sigma D_\sigma \bigl(e \partial_x V_\sigma - \sigma \dot Z_x^{so}\bigr).
\label{current_normal_2}
\end{eqnarray}

Assuming that the interlayer is a half-metal, that is the density of states is nonzero only for one of the spin subbands, we obtain from Eq.~(\ref{current_normal_2}) that the potential gradient in the interlayer takes the form:
\begin{eqnarray}
e \partial_x V = -\frac{j}{e \nu D} + \dot Z_x^{so}.
\label{V_derivative}
\end{eqnarray}

Let us choose the electric potential in the left electrode to be zero $V_L = 0$. Then the electric potential of the right electrode 
\begin{equation}
V_R = V_{L,b} + V_{int} + V_{R,b},
\label{potential_drop}
\end{equation} 

where $V_{L(R),b} = -jR_{L(R)}$ is the potential jump at the left (right) S/F interface and $V_{int} = \int \limits_{-d/2}^{d/2} \partial_x V dx$ is the potential difference acquired inside the interlayer.

Substituting Eq.~(\ref{V_derivative}) into Eq.~(\ref{potential_drop}) we obtain
\begin{eqnarray}
j=\frac{-(V_R - V_L) + (1/e)\int \limits_{-d/2}^{d/2} \dot Z_x^{so}dx}{R} ,
\label{current_final}
\end{eqnarray}
where we have introduced the total normal resistance of the junction $R = R_L + R_R + d/(e^2 \nu D)$.

Recalling that $V_L - V_R = \dot \varphi/2e$ and $\dot \varphi_0 = -2 \int \limits_{-d/2}^{d/2} \dot Z_x^{so}dx$ we finally obtain
\begin{equation}
j_n = \frac{\dot \varphi - \dot \varphi_0}{2eR}.
\end{equation}

\subsection*{DW motion induced by a given current profile}

Here we derive exact expressions for the DW velocity as a function of time for different current regimes considered in the main text. The full dependence $v(t)$ can be obtained starting from the LLG equation (\ref{LLG}). Substituting $\cos \theta$ in the form of Eq.~(\ref{theta}) and $\delta = \delta_0 + \Delta \delta (t)$, where $\delta_0 = \pi/2$ for a Neel DW, into Eq.~(\ref{LLG}), assuming $\Delta \delta \ll 1 $ and keeping only first order terms with respect to this parameter, after some algebra we obtain the following expression for $v(t)$:
\begin{eqnarray}
v(t)=\exp(-t/t_d) \times \nonumber\\
\int_{-0}^{t}\exp(t'/t_d)
\left(-u(t')\dfrac{\beta}{\alpha t_d}+{u}'(t')\dfrac{1-\alpha\beta}{1+\alpha^2}\right)dt'\label{v(t)},~~~~
\label{velocity_general}
\end{eqnarray}
where $t_d = (1+\alpha^2)M/\alpha \gamma K_\perp$ and $u(t) = \gamma j(t)/2eM$. Eq.~(\ref{velocity_general}) is valid for an arbitrary dependence of $j(t)$ if the current is switched on at $t=0$.

If a constant current $j(t) = j\theta(t)$ is switched on at $t=0$, then one can obtain from Eq.~(\ref{velocity_general}) that at $t>0$
\begin{eqnarray}
v(t)=\frac{e^{-t/t_d} u}{1+\alpha^2}\left(1+\frac{\beta}{\alpha}\right)-\frac{\beta u}{\alpha}.
\label{velocity_constant}
\end{eqnarray}
As it was stated in the main text, the DW velocity at $t>0$ saturates exponentially to the value $v_{st}$ defined by Eq.~(\ref{velocity_steady}). 

For the case of a rectangular current impulse $j(t)=j\theta(t)\theta(T-t)$ the DW velocity takes the form:
\begin{eqnarray}
	v(t)=-\exp(-t/t_d)\theta(t)\dfrac{\gamma j}{2eM}\left[\dfrac{\alpha+\beta}{\alpha(1+\alpha^2)}\times\right.\nonumber~~~~~~~~~~~\\
	\left.\times\Bigl(\exp(T/t_d)\theta(t-T)-1\Bigr)+\dfrac{\beta}{\alpha}\exp(t/t_d)\theta(T-t)\right]~~~~~~\label{v_delta}
\end{eqnarray}

\subsection*{Voltage induced by a rectangular current impulse}

Having at hand the dependence $v(t)$ one can find the anomalous phase shift as a function of time. According to Eq.~(\ref{varphi_0_approx}) of the main text
\begin{eqnarray}
	\varphi_0(t) \approx -2 \pi \beta/d_W \int \limits_0^t v(t')dt'.
	\label{varphi_0_velocity}
\end{eqnarray}

Solving Eq.~(\ref{josephson_modified}) with $j(t)=j\theta(t)\theta(T-t)$ and $\varphi_0(t)$ defined by Eqs.~(\ref{varphi_0_velocity}) and (\ref{v_delta}) we obtain:
\begin{eqnarray}
\dot \varphi(t)=\dfrac{\Omega (\frac{j^2}{j_c^2}-1)\theta(T-t)}{\frac{j}{j_c}-\cos(\Omega t \sqrt{(j/j_c)^2-1}+\arctan\sqrt{(j/j_c)^2-1})} \nonumber\\
-\dfrac{2\Omega D \exp(-\Omega t)\theta(t-T)}{1+D^2\exp(-2\Omega t)}-\frac{2\pi \beta v(t)}{d_W},~~~~~~~~~~~~~`
\label{voltage_impulse}
\end{eqnarray}
where 
$D=(j_c/j)\exp(\Omega T)(1-\sqrt{(j/j_c)^2-1}\times\\ \times\cot(\Omega T\sqrt{(j/j_c)^2-1}/2+\arctan\sqrt{(j/j_c)^2-1}))$, $\Omega =t_J^{-1}= 2e RI_c$. At $j<j_c$   $\sqrt{(j/j_c)^2-1} \to i\sqrt{1-(j/j_c)^2}$.
$V(t)$ described by Eq.~(\ref{voltage_impulse}) is plotted in Fig.~3 of the main text.


\begin{thebibliography}{1}%
\makeatletter
\providecommand \@ifxundefined [1]{%
 \@ifx{#1\undefined}
}%
\providecommand \@ifnum [1]{%
 \ifnum #1\expandafter \@firstoftwo
 \else \expandafter \@secondoftwo
 \fi
}%
\providecommand \@ifx [1]{%
 \ifx #1\expandafter \@firstoftwo
 \else \expandafter \@secondoftwo
 \fi
}%
\providecommand \natexlab [1]{#1}%
\providecommand \enquote  [1]{``#1''}%
\providecommand \bibnamefont  [1]{#1}%
\providecommand \bibfnamefont [1]{#1}%
\providecommand \citenamefont [1]{#1}%
\providecommand \href@noop [0]{\@secondoftwo}%
\providecommand \href [0]{\begingroup \@sanitize@url \@href}%
\providecommand \@href[1]{\@@startlink{#1}\@@href}%
\providecommand \@@href[1]{\endgroup#1\@@endlink}%
\providecommand \@sanitize@url [0]{\catcode `\\12\catcode `\$12\catcode
  `\&12\catcode `\#12\catcode `\^12\catcode `\_12\catcode `\%12\relax}%
\providecommand \@@startlink[1]{}%
\providecommand \@@endlink[0]{}%
\providecommand \url  [0]{\begingroup\@sanitize@url \@url }%
\providecommand \@url [1]{\endgroup\@href {#1}{\urlprefix }}%
\providecommand \urlprefix  [0]{URL }%
\providecommand \Eprint [0]{\href }%
\providecommand \doibase [0]{http://dx.doi.org/}%
\providecommand \selectlanguage [0]{\@gobble}%
\providecommand \bibinfo  [0]{\@secondoftwo}%
\providecommand \bibfield  [0]{\@secondoftwo}%
\providecommand \translation [1]{[#1]}%
\providecommand \BibitemOpen [0]{}%
\providecommand \bibitemStop [0]{}%
\providecommand \bibitemNoStop [0]{.\EOS\space}%
\providecommand \EOS [0]{\spacefactor3000\relax}%
\providecommand \BibitemShut  [1]{\csname bibitem#1\endcsname}%
\let\auto@bib@innerbib\@empty
\bibitem [{Note1()}]{Note1}%
  \BibitemOpen
  \bibinfo {note} {It is worth to mention that for an in-plane exchange field
  as is shown in Fig.~\ref {sketch}(b) the Rashba SOC by itself does not
  generate long-range triplet correlations at S/F interfaces. Therefore, to
  produce them spin-active layers should be added to the system at the S/F
  interfaces. However, this problem is widely studied in the literature\cite
  {Bergeret2005,Eschrig2008,Houzet2007} and is not discussed here.}\BibitemShut
  {Stop}%
\end{thebibliography}%


\begin{thebibliography}{99}
%
\bibitem{Abrikosov1957}
A. A. Abrikosov, Zh. Eksperim. i Teoret. Fiz. {\bf 32}, 1442 (1957) [
Soviet Phys. JETP {\bf 6}, 1174 (1957)].
%
\bibitem{Kim1965}
Y. B. Kim, C. F. Hempstead, and A. R. Strnad, Phys. Rev. {\bf 139}, A1163 (1965).
%
\bibitem{Bardeen1965}
J. Bardeen and M. J. Stephen, Phys. Rev. {\bf 140}, A1197 (1965).
%
\bibitem{Gorkov1975}
L. P. Gor’kov and N. B. Kopnin, Soviet Physics
Uspekhi {\bf 18}, 496 (1975).
%
\bibitem{Bergeret2005}
F. S. Bergeret, A. F. Volkov, K. B. Efetov, Rev. Mod. Phys.
{\bf 77}, 1321 (2005).
%
\bibitem{BuzdinRMP2005}
A. I. Buzdin , 
Rev. Mod. Phys. {\bf 77}, 935 (2005)
%
\bibitem{BergeretRMP2018}
F. S. Bergeret, M. Silaev, P. Virtanen, and T. T. Heikkila
Rev. Mod. Phys. {\bf 90}, 041001 (2018) 
%
\bibitem{Eschrig2008}
M. Eschrig, T. Lofwander, Nat. Phys. {\bf 4}, 138 (2008).
%
\bibitem{Houzet2007}
 M. Houzet, A. I. Buzdin, Phys. Rev. B {\bf 76}, 060504(R)
(2007).
%
\bibitem{Eschrig2015}
M. Eschrig,  Rep.Prog. Phys. {\bf 78},  104501 (2015) 
%
\bibitem{Waintal2002}
X. Waintal and P. W. Brouwer, Phys. Rev. B {\bf 65}, 054407 (2002).
%

\bibitem{Buzdin2008}
A.I. Buzdin, Phys. Rev. Lett. {\bf 101}, 107005 (2008).
\bibitem{Konshelle2009}
F. Konschelle, A. Buzdin
Phys. Rev. Lett. {\bf 102}, 017001 (2009)
%
\bibitem{Teber2010}
S. Teber, C. Holmqvist, and M. Fogelstrom,
Phys. Rev. B {\bf 81}, 174503 (2010)
%
\bibitem{Holmqvist2012}
C. Holmqvist, W. Belzig, and M. Fogelstrom,
Phys. Rev. B {\bf 86}, 054519 (2012)
%
\bibitem{Linder2011}
J. Linder and T. Yokoyama, Phys. Rev. B {\bf 83}, 012501 (2011).
%
\bibitem{Shukrinov2017}
Yu. M. Shukrinov, I. R. Rahmonov, K. Sengupta, and A. Buzdin, Appl. Phys. Lett. {\bf 110}, 182407 (2017).
%
\bibitem{Chudnovsky2016}
E. M. Chudnovsky, Phys. Rev. B {\bf 93}, 144422 (2016).
%

\bibitem{Braude2008}
V. Braude and Ya. M. Blanter, Phys. Rev. Lett. {\bf 100}, 207001 (2008).

%
\bibitem{Nussinov2005}
Z. Nussinov, A. Shnirman, D. P. Arovas, A. V. Balatsky, and J.-X. Zhu,
Phys. Rev. B {\bf 71}, 214520 (2005)
%
\bibitem{Zhu2004}
J.-X. Zhu, Z. Nussinov, A. Shnirman, and A. V. Balatsky,
Phys. Rev. Lett. {\bf 92}, 107001 (2004)


\bibitem{Cai2010}
L. Cai and E. M. Chudnovsky, Phys. Rev. B {\bf 82}, 104429 (2010).
%
\bibitem{Bobkova2018}
I. V. Bobkova, A. M. Bobkov, and M.A. Silaev, Phys. Rev. B {\bf 98}, 014521 (2018).
%
\bibitem{Rabinovich2018}
D.S. Rabinovich, I.V. Bobkova, A.M. Bobkov and M.A. Silaev, Phys. Rev. B {\bf 98}, 184511 (2018). 
%
\bibitem{Aikebaier2019}
F. Aikebaier, P. Virtanen, and T.T. Heikkila,
Phys. Rev. B {\bf 99}, 104504 (2019)
%

\bibitem{ParkinRacetrack1}
 S. S. P. Parkin, M. Hayashi, and L. Thomas, Science {\bf 320}, 190
(2008).

\bibitem{ParkinRacetrack2}
M. Hayashi, L. Thomas, R. Moriya, C. Rettner, and S. S. P.
Parkin, Science {\bf 320}, 209 (2008).

\bibitem{Singh2016}
A. Singh, C. Jansen, K. Lahabi, and J. Aarts, Phys. Rev.
X {\bf 6}, 041012 (2016).

\bibitem{Bobkova2017}
I.V. Bobkova, A.M. Bobkov, and M.A. Silaev, Phys. Rev. B {\bf 96}, 094506 (2017).
%
\bibitem{Frohlich1993}
 J. Frohlich and U. M. Studer, Rev. Mod. Phys. {\bf 65}, 733
(1993).
%
\bibitem{Rebei2006}
A. Rebei and O. Heinonen, Phys. Rev. B {\bf 73}, 153306
(2006).
%
\bibitem{Jin2006}
P.-Q. Jin, Y.-Q. Li, and F.-C. Zhang, J. Phys. A
{\bf 39}, 7115 (2006).
%
\bibitem{Jin2006_2}
P.-Q. Jin and Y.-Q. Li, Phys. Rev. B {\bf 74},
085315 (2006).
%
\bibitem{Bernevig2006}
B. A. Bernevig, J. Orenstein, and S.-C.
Zhang, Phys. Rev. Lett. {\bf 97}, 236601 (2006).
%
\bibitem{Hatano2007}
N. Hatano, R. Shirasaki, and H. Nakamura, Phys. Rev. A {\bf 75}, 032107
(2007).
%
\bibitem{Leurs2009}
B. W. A. Leurs, Z. Nazario, D. I. Santiago, and J. Zaanen, Annals of Physics, {\bf 324}, 1821 (2009)
%
\bibitem{Tokatly2008}
I. V. Tokatly, Phys. Rev. Lett. {\bf 101}, 106601 (2008).
%
\bibitem{Bergeret2013}
F. S. Bergeret and I. V. Tokatly, Phys. Rev. Lett. {\bf 110}, 117003 (2013).
%
%
\bibitem{Kim2012}
K. W. Kim, J. H. Moon, K. J. Lee, and H. W. Lee, Phys. Rev.
Lett. {\bf 108}, 217202 (2012).
%
\bibitem{Tatara2013}
G. Tatara, N. Nakabayashi, and K. J. Lee, Phys. Rev. B {\bf 87}, 054403 (2013).
%
\bibitem{Yamane2013}
Y. Yamane, J. Ieda, and S. Maekawa, Phys. Rev. B {\bf 88}, 014430 (2013).
%
\bibitem{Slonczewski1996}
J. C. Slonczewski, Journal of Magnetism and Magnetic Materials
{\bf 159}, L1 (1996).
%
\bibitem{Tatara2004}
G. Tatara and H. Kohno, Phys. Rev. Lett. {\bf 92}, 086601 (2004).
%
\bibitem{Koyama2011}
T. Koyama, D. Chiba, K. Ueda, K. Kondou, H. Tanigawa, S. Fukami, T. Suzuki, N. Ohshima, N. Ishiwata, Y. Nakatani, K. Kobayashi, and T. Ono, Nature Materials
{\bf 10}, 194 (2011).
%
\bibitem{Miron2010}
I. M. Miron, G. Gaudin, S. Auffret, B. Rodmacq, A. Schuhl, S. Pizzini, J. Vogel, and P. Gambardella, Nature Mater. {\bf 9}, 230 (2010).
%
\bibitem{Gambardella2011}
P. Gambardella and I. M. Miron, Philos Transact A Math Phys Eng Sci {\bf 369}, 3175 (2011).
%

\bibitem{Zhang2004}
S. Zhang and Z. Li, Phys. Rev. Lett. {\bf 93}, 127204 (2004).
%
\bibitem{Asano2007}
Y. Asano, Y. Sawa, Y. Tanaka, and A.A. Golubov, Phys. Rev. B {\bf 76}, 224525 (2007).
%
\bibitem{Reynoso2008}
A.A. Reynoso, G. Usaj, C.A. Balseiro, D. Feinberg, and M. Avignon, Phys. Rev. Lett. {\bf 101}, 107001 (2008).
%
\bibitem{Tanaka2009}
Y. Tanaka, T. Yokoyama and N. Nagaosa, Phys. Rev. Lett. {\bf 103}, 107002 (2009).
%
\bibitem{Grein2009}
R. Grein, M. Eschrig, G. Metalidis, and G. Schon, Phys. Rev. Lett. {\bf 102}, 227005 (2009).
%
\bibitem{Zazunov2009}
A. Zazunov, R. Egger, T. Jonckheere, and T. Martin, Phys. Rev. Lett. {\bf 103}, 147004 (2009).
%
\bibitem{Liu2010}
J.-F. Liu, K. S. Chan, Phys. Rev. B {\bf 82}, 184533 (2010).
%
\bibitem{Malshukov2010}
A.G. Mal’shukov, S. Sadjina, and A. Brataas, Phys. Rev. B {\bf 81}, 060502 (2010).
%
\bibitem{Alidoust2013}
M. Alidoust and J. Linder, Phys. Rev. B {\bf 87}, 060503(R) (2013).
%
\bibitem{Brunetti2013}
A. Brunetti, A. Zazunov, A. Kundu, and R. Egger Phys. Rev. B {\bf 88}, 144515 (2013).
%
\bibitem{Yokoyama2014}
T. Yokoyama, M. Eto, and Yu. V. Nazarov, Phys. Rev. B {\bf 89}, 195407 (2014).
%
\bibitem{Kulagina2014}
I. Kulagina and J. Linder, Phys. Rev. B {\bf 90}, 054504 (2014).
%
\bibitem{Moor2015_1}
A. Moor, A.F. Volkov, and K.B. Efetov, Phys. Rev. B {\bf 92}, 214510 (2015).
%
\bibitem{Moor2015_2}
A. Moor, A. F. Volkov, and K. B. Efetov, Phys. Rev. B {\bf 92}, 180506 (2015).
%
\bibitem{Bergeret2015}
F. S. Bergeret and I. V. Tokatly, EPL {\bf 110}, 57005 (2015).
%
\bibitem{Campagnano2015}
G. Campagnano, P. Lucignano, D. Giuliano and A. Tagliacozzo, J. Phys. Cond. Mat. {\bf 27}, 205301 (2015).
%
\bibitem{Mironov2015}
S. Mironov and A. Buzdin, Phys. Rev. B {\bf 92}, 184506 (2015).
%
\bibitem{Konschelle2015}
F. Konschelle, I. V. Tokatly, and F. S. Bergeret, Phys.
Rev. B {\bf 92}, 125443 (2015).
%
\bibitem{Kuzmanovski2016}
D. Kuzmanovski, J. Linder, A. Black-Schaffer, Phys. Rev. B {\bf 94}, 180505 (2016).
%
\bibitem{Zyuzin2016} 
A. A. Zyuzin, M. Alidoust, and D. Loss, {Phys. Rev. B {\bf 93}, 214502 (2016)}.
%
\bibitem{Bobkova2016}
I. V. Bobkova, A. M. Bobkov, A. A. Zyuzin, and M. Alidoust, Phys. Rev. B {\bf 94}, 134506 (2016).
%
\bibitem{Silaev2017}
M. A. Silaev, I. V. Tokatly, and F. S. Bergeret, Phys. Rev.
B {\bf 95}, 184508 (2017).
%
\bibitem{Szombati2016}
D.B. Szombati,  S. Nadj-Perge, D. Car, S. R. Plissard, E. P. A. M. Bakkers, and L. P. Kouwenhoven, Nature Phys. {\bf 12}, 568 (2016). 
%
\bibitem{Murani2017}
A. Murani,  A. Kasumov, S. Sengupta, Yu. A. Kasumov, V.T. Volkov,  I. I. Khodos, F. Brisset,  R. Delagrange, A. Chepelianskii, R. Deblock, H. Bouchiat, and S. Guéron, Nature Comm. {\bf 8}, 15941 (2017). 
%
\bibitem{Assouline2019}
A. Assouline, C. Feuillet-Palma, N. Bergeal, T. Zhang, A. Mottaghizadeh, A. Zimmers, E. Lhuillier, M. Marangolo, M. Eddrief, P. Atkinson, M. Aprili, and H. Aubin, Nat. Commun. {\bf 10}, 126 (2019).
%
\bibitem{Meng2019}
H. Meng, A. V. Samokhvalov, and A. I. Buzdin, PRB {\bf 99}, 024503 (2019)
%
\bibitem{Krive2004}
I.V. Krive, L.Y. Gorelik, R.I. Shekhter, and M. Jonson, Phys. Nizk. Temp. {\bf 30}, 535 (2004).
%
\bibitem{Braude2007}
V. Braude and Yu.V. Nazarov, Phys. Rev. Lett. {\bf 98}, 077003 (2007).
%
\bibitem{Volovik1987}
G. E. Volovik, J. Phys. C {\bf 20}, L83 (1987).
%
\bibitem{Barnes2007}
S.E. Barnes and S. Maekawa, Phys. Rev. Lett. {\bf 98}, 246601 (2007).
%
\bibitem{Saslow2007}
W. M. Saslow, Phys. Rev. B {\bf 76}, 184434 (2007).
%
\bibitem{Duine2008}
R. A. Duine, Phys. Rev. B {\bf 77}, 014409 (2008).
%
\bibitem{Tserkovnyak2008}
Y. Tserkovnyak and M. Mecklenburg, Phys. Rev. B {\bf 77}, 134407 (2008).
%
\bibitem{Zhang2009}
S. Zhang and S. S.-L. Zhang, Phys. Rev. Lett. {\bf 102}, 086601 (2009).

\bibitem{Schulz2012}
T. Schulz et al.,  Nature Physics {\bf 8}, 301 (2012)
%
\bibitem{Nagaosa2013}
N. Nagaosa, Y. Tokura, Nature Nanotechnology {\bf 8}, 899 (2013)
%
\bibitem{Stewart1968}
W. C. Stewart, Appl. Phys. Lett. {\bf 12}, 277 (1968). 
%
\bibitem{McCumber1968}
D. E. McCumber, J. Appl. Phys. {\bf 39}, 3113 (1968).
%
\bibitem{Kautz1990}
R. L. Kautz and John M. Martinis Phys. Rev. B {\bf 42}, 9903 (1990).
%
\bibitem{Golod2019}
T. Golod, O.M. Kapran, and V.M. Krasnov, Phys. Rev. Applied {\bf 11}, 014062 (2019). 
%
\bibitem{LJJRatchet1}
 E. Goldobin, A. Sterck, and D. Koelle, Phys. Rev. E 63,
031111 (2001); G. Carapella, Phys. Rev. B {\bf 63}, 054515
(2001); A. O. Sboychakov, S. Savel’ev, A. L. Rakhmanov,
and F. Nori, Phys. Rev. Lett. {\bf 104}, 190602 (2010).
%
\bibitem{LJJRatchet2}
 G. Carapella and G. Costabile, Phys. Rev. Lett. 87, 077002
(2001).

\bibitem{LJJRatchet3}
 M. Beck, E. Goldobin, M. Neuhaus, M. Siegel, R. Kleiner,
and D. Koelle, Phys. Rev. Lett. {\bf 95}, 090603 (2005).

\bibitem{JJarrayRatchet}
 E. Trias, J. J. Mazo, F. Falo, and T. P. Orlando, Phys. Rev. E {\bf 61}, 2257 (2000); 
F. Falo, P. J. Martinez, J. J. Mazo, and
S. Cilla, Europhys. Lett. {\bf 45}, 700 (1999); 
K. H. Lee, Appl.
Phys. Lett. {\bf 83}, 117 (2003); V. I. Marconi, Phys. Rev. Lett.
{\bf 98}, 047006 (2007).

\bibitem{asymSQUIDRatchet1}
I. Zapata, R. Bartussek, F. Sols, and P. Hanggi, Phys. Rev.
Lett. {\bf 77}, 2292 (1996).

\bibitem{asymSQUIDRatchet2}
S. Weiss, D. Koelle, J. Muller, R. Gross, and K. Barthel,
Europhys. Lett. {\bf 51}, 499 (2000).

\bibitem{3JSQUIDRatchet1}
A. Sterck, R. Kleiner, and D. Koelle, Phys. Rev. Lett. {\bf 95},
177006 (2005).

\bibitem{3JSQUIDRatchet2}
 A. Sterck, D. Koelle, and R. Kleiner, Phys. Rev. Lett. {\bf 103}, 047001 (2009).
 
 \bibitem{AnnularJJRatchet}
 A. Davidson, B. Dueholm, B. Kryger, and N. F. Pedersen,
Phys. Rev. Lett. {\bf 55}, 2059 (1985); 
A. V. Ustinov, T.
Doderer, R. P. Huebener, N. F. Pedersen, B. Mayer, and
V. A. Oboznov, Phys. Rev. Lett. {\bf 69}, 1815 (1992); 
N.
Gronbech-Jensen, P. S. Lomdahl, and M. R. Samuelsen,
Phys. Lett. A {\bf 154}, 14 (1991); 
A. Wallraff, Yu. Koval, M.
Levitchev, M. V. Fistul, and A. V. Ustinov, J. Low Temp.
Phys. {\bf 118}, 543 (2000). 

 \bibitem{asymDriveJJRatchet1}
 A. V. Ustinov, C. Coqui, A. Kemp, Y. Zolotaryuk, and M.
Salerno, Phys. Rev. Lett. {\bf 93}, 087001 (2004).

 \bibitem{asymDriveJJRatchet2}
F. Marchesoni, Phys. Lett. A 
{\bf 119}, 221 (1986); S. Flach, Y.
Zolotaryuk, A. E. Miroshnichenko, and M. V. Fistul, 
Phys.
Rev. Lett. {\bf 88}, 184101 (2002).

 \bibitem{AbrikosovVortJJRatchet}
 G. R. Berdiyorov, M. V. Milosevic, L. Covaci, and F. M. Peeters, Phys.
Rev. Lett. {\bf 107}, 177008 (2011).
%
\bibitem{Ast2007}
C. R. Ast, J. Henk, A. Ernst, L. Moreschini, M. C. Falub, D. Pacile, P. Bruno, K. Kern, and M. Grioni, Phys. Rev. Lett. {\bf 98}, 186807 (2007).
%
\bibitem{Tserkovnyak2009}
Y. Tserkovnyak and C. H. Wong, Phys. Rev. B {\bf 79}, 014402 (2009).
%
\bibitem{Shapiro1963}
 S. Shapiro, Phys. Rev. Lett. {\bf 11}, 80 (1963). 
%
\bibitem{Grimes1968}
C. C. Grimes and S. Shapiro, Phys. Rev. {\bf 169}, 397 (1968).
%
\bibitem{Khaire2010}
T.S. Khaire, M. A. Khasawneh, W. P. Pratt, Jr., and N. O. Birge, Phys. Rev. Lett. {\bf 104}, 137002 (2010). 
%
\bibitem{Robinson2010}
J. W. A. Robinson, J. D. S. Witt, M. G. Blamire, Science {\bf 329}, 59 (2010).
%
\bibitem{Zou2007}
X. Zou and G. Xiao, Appl. Phys. Lett. {\bf 91}, 113512 (2007).
%







\end{thebibliography}
\end{document}